\documentclass{article}
\usepackage[numbers]{natbib}
\usepackage[margin=1in]{geometry}

\newenvironment{affiliations}{\medskip\large}

\newcommand{\keywords}[1]{\medskip Keywords: \textit{#1}}

\usepackage[version=4]{mhchem}

\usepackage{tikz}
\usetikzlibrary{shapes.geometric, arrows.meta, positioning}
\usepackage{booktabs}
\usepackage{multirow}
\usepackage{siunitx}
\usepackage{graphicx}
\usepackage{subcaption}

\usepackage{titlesec}

\titlespacing{\paragraph}{%
    0pt}{
    0pt}{
    1em}

\usepackage{soul}

\definecolor{mpink}{HTML}{BF016B}

\usepackage[acronym]{glossaries}
\glsdisablehyper  
\makeglossaries
\newacronym{ace}{ACE}{Atomic Cluster Expansion}
\newacronym{acsf}{ACSF}{Atom-centered symmetry functions}
\newacronym{aimd}{AIMD}{\textit{ab initio} molecular dynamics}
\newacronym{ccdc}{CCDC}{Cambridge Structural Database}
\newacronym{csd}{CSD}{Cambridge Structure Database}
\newacronym{md}{MD}{Molecular Dynamics}
\newacronym{bob}{BoBs}{Bags of Bonds}
\newacronym{dft}{DFT}{Density Functional Theory}
\newacronym{dof}{DOF}{Degrees of Freedom}
\newacronym{gap}{GAP}{Gaussian Approximation Potentials}
\newacronym{snap}{SNAP}{Spectral Neighbor Analysis Potential} 
\newacronym{gn}{GN}{Graph Networks}
\newacronym{gdml}{GDML}{Gradient-Domain Machine Learning}
\newacronym{hd}{HD}{High-Dimensional}
\newacronym{krr}{KRR}{Kernel Ridge Regression}
\newacronym{mc}{MC}{Monte Carlo}
\newacronym{gen}{Gen}{Generation}
\newacronym{ld}{LD}{Low-Dimensional}
\newacronym{lri}{LRI}{Long-Range Interactions}
\newacronym{nli}{NLI}{non-local interaction}
\newacronym{m-s}{M-S}{molecule-surface}
\newacronym{ml}{ML}{Machine Learning}
\newacronym{rdf}{RDF}{Radial Distribution function}
\newacronym{difftre}{DiffTRe}{Differentiable Trajectory Reweighting}
\newacronym{diffmd}{DiffMD}{Diffusion Model for Molecular Dynamics}
\newacronym{ad}{AD}{Automatic Differentiation}
\newacronym{ide}{IDE}{integrated development environment}
\newacronym{lj}{LJ}{Lennard-Jones}
\newacronym{eam}{EAM}{Embedded Atom Model}
\newacronym{ks-dft}{KS-DFT}{ Kohn--Sham Density Functional Theory}
\newacronym{scf}{SCF}{Scaling Correlation Factor}
\newacronym{scan}{SCAN}{Strongly Constrained and Appropriately Normed}
\newacronym{gnn}{GNN}{ graph neural network}
\newacronym{stable}{StABlE}{Stability-Aware-Boltzmann Estimator}
\newacronym{mlip}{MLIP}{Machine Learning Interaction Potential}
\newacronym{mtp}{MTP}{Moment Tensor Potential}
\newacronym{nnp}{NNP}{Neural Network Potential}
\newacronym{nn}{NN}{Neural Network}
\newacronym{pes}{PES}{Potential Energy Surface}
\newacronym{pip}{PIP}{Permutation Invariant Polynomials}
\newacronym{piv}{PIV}{Permutation Invariant Vector}
\newacronym{qm}{QM}{Quantum Mechanics}
\newacronym{soap}{SOAP}{Smooth Overlap of Atomic Positions}
\newacronym{sprint}{SPRINT}{Social PermuASE WF for the training of a Machine Learning Interaction Potential(MLIP)),tation Invariant Topological coordinates}
\newacronym{sri}{SRI}{Short-Range Interactions}
\newacronym{svm}{SVM}{Support Vector Machines}
\newacronym{ann}{ANN}{Artificial Neuronal Network}
\newacronym{ffnn}{FFNN}{Feed-Forward Neuronal Network}
\newacronym{mffnn}{MFFNN}{Multi-layers Feed Forward Neural Network}
\newacronym{hdmr}{HDMR}{High Dimensional Model Representation}
\newacronym{hdnnp}{HDNNP}{High-Dimensional Neural Network Potential}
\newacronym{expnn}{ExpNN}{Exponential Neural Network}
\newacronym{rc/rs/hdmr}{RC-RS-HDMR}{ Redundant Coordinates Random Sampling High Dimensional Model Representation }
\newacronym{pe}{PE}{Polyethylene}
\newacronym{c}{C}{Carbon}
\newacronym{ref}{Ref}{Reference}
\newacronym{bh4}{BH4}{Tetrahydrobiopterin}
\newacronym{bpnn}{BPNN}{Back-Propagation Neural Network}
\newacronym{snnp}{SNNP}{Symmetry Adapted Neural Network Potential}
\newacronym[longplural=self-driving laboratories]{sdl}{SDL}{self-driving laboratory}
\newacronym{ens/ffnn}{EnsFFNNs}{Ensemble Feed-Forward Neuronal Networks}
\newacronym{asnn}{ASNN}{Associative Neuronal Networks}
\newacronym{rbfnn}{RBFNN}{Radial Basis Function Neural Network}
\newacronym{gpr}{GPR}{Gaussian Process Regression}
\newacronym{dma}{DMA}{Dimethylacetamide}
\newacronym{inf}{\infty}{Dimensional Complexity}
\newacronym{ug}{UG}{Uncertainty Quantification }
\newacronym{wf}{WF}{Workflow}
\newacronym{icsd}{ICSD}{Inorganic Crystal Structure Database}
\newacronym{asm}{ASM}{American Society for Metals}
\newacronym{calphad}{CALPHAD}{Calculation of Phase diagram}
\newacronym{hpc}{HPC}{High-Performance Computing}
\newacronym{vm}{VM}{virtual machines}
\newacronym{bpnnmodel}{BPNN}{Behler-Parrinello neuronal network}
\newacronym{bcdi}{BCDI}{Bragg coherent diffraction imaging}
\newacronym{wms}{WMS}{workflow management systems}

\usepackage{listings}
\usepackage{xcolor}
\begin{document}


\title{Reproducible container solutions for codes and workflows in materials science}

\maketitle

\author{Dylan Bissuel,}
\author{Léo Orveillon,}
\author{Benjamin Arrondeau,}
\author{João Paulo Almeida de Mendonça,}
\author{Irina Piazza,}
\author{Martin Uhrin,}
\author{Étienne Polack,}
\author{Akshay Krishna Ammothum Kandy,}
\author{David Martin-Calle,}
\author{Jonathan Chapignac,}
\author{Aadhityan Arivazhagan,}
\author{Lorenzo Paulatto,}
\author{Pierre-Antoine Bouttier,}
\author{M.-I. Richard,}
\author{Thierry Deutsch,}
\author{David Rodney,}
\author{A. M. Saitta,}
\author{N\"{o}el Jakse*}


\begin{affiliations}

  D. Bissuel\\
  Université Claude Bernard Lyon 1, CNRS, Institut Lumière Matière, F-69622 Villeurbanne, France\\

  L. Orveillon\\
  GRICAD, Université Grenoble Alpes, F-38400 Grenoble, France\\

  B. Arrondeau\\
  GRICAD, Université Grenoble Alpes, F-38400 Grenoble, France\\

  J.P. Almeida de Mendonça\\
  Université Grenoble Alpes, CNRS, Grenoble INP, SIMaP, F-38000 Grenoble, France\\

  I. Piazza\\
  Université Grenoble Alpes, CNRS, Grenoble INP, SIMaP, F-38000 Grenoble, France\\
  Email: irina.piazza@grenoble-inp.fr\\

  M. Uhrin\\
  Université Grenoble Alpes, CNRS, Grenoble INP, SIMaP, F-38000 Grenoble, France\\

  É. Polack\\
  Université Grenoble Alpes, CEA, IRIG, MEM, NRX, F-38000 Grenoble, France\\

  A. K. A. Kandy\\
  Sorbonne Université, MNHN, CNRS, Institut de Minéralogie, de Physique des Matériaux et de Cosmochimie, F-75005 Paris, France\\

  D. Martin-Calle\\
  Université Claude Bernard Lyon 1, CNRS, Institut Lumière Matière, F-69622 Villeurbanne, France\\

  J. Chapignac\\
  Université Grenoble Alpes, CNRS, Grenoble INP, SIMaP, F-38000 Grenoble, France\\

  A. Arivazhagan\\
  Sorbonne Université, MNHN, CNRS, Institut de Minéralogie, de Physique des Matériaux et de Cosmochimie, F-75005 Paris, France\\

  L. Paulatto\\
  Sorbonne Université, MNHN, CNRS, Institut de Minéralogie, de Physique des Matériaux et de Cosmochimie, F-75005 Paris, France\\

  P.-A. Bouttier\\
  GRICAD, Université Grenoble Alpes, F-38400 Grenoble, France\\

  M.-I. Richard\\
  Université Grenoble Alpes, CEA, IRIG, MEM, NRX, F-38000 Grenoble, France\\

  T. Deutsch\\
  Université Grenoble Alpes, CEA, IRIG, MEM, F-38000 Grenoble, France\\

  D. Rodney\\
  Université Claude Bernard Lyon 1, CNRS, Institut Lumière Matière, F-69622 Villeurbanne, France\\

  A. M. Saitta\\
  École Normale Supérieure--PSL, CNRS, Sorbonne Université, Université Paris-Cité, Laboratoire de Physique de l'ENS, F-75005 Paris, France\\

  N. Jakse\\
  Université Grenoble Alpes, CNRS, Grenoble INP, SIMaP, F-38000 Grenoble, France\\

\end{affiliations}



\begin{abstract}
  A computing solution combining the GNU Guix functional package manager with the Apptainer container system is presented. This approach provides fully declarative and reproducible software environments suitable for computational materials science. Its versatility and performance enable the construction of complete frameworks integrating workflow managers such as AiiDA, and Ewoks that can be deployed on HPC infrastructures. The efficiency of the solution is illustrated through several examples: (i) AiiDA workflows for automated dataset construction and analysis as well as path-integral molecular dynamics based on \textit{ab initio} calculations; (ii) workflows for the training of machine-learning interatomic potentials; and (iii) an Ewoks workflow for the automated analysis of coherent X-ray diffraction data in large-scale synchrotron facilities. These examples demonstrate that the proposed environment provides a reliable and reproducible basis for computational and data-driven research in materials science.

\end{abstract}

\section{Introduction}\label{sec1}

The combination of a vast amount of accumulated data and the continuous increase in computing power has led to the rapid development of \gls{ml} methods in materials science~\cite{schmidt2019recent,SCHMIDT2024101560}. These methods have not only accelerated research but also enabled the identification of solutions to complex problems that were previously intractable. The predictive ability of \gls{ml} models, however, inherently depends on the quality and quantity of data used during training and validation~\cite{rodrigues2021big}.

Building a \gls{ml} model typically begins with the collection of data from experiments or computations, often gathered from accessible databases.
The construction of internationally available databases began in the 1960s with crystallographic repositories based on manually curated experimental crystal structures, such as the \gls{icsd} and \gls{csd} for inorganic and organic molecules, respectively~\cite{gorbitz2016development}. Later, in the 1980s, as computing power increased, structured databases emerged for alloy phase diagrams, such as the Alloy Phase Diagram Database from \gls{asm}~\cite{asm_alloy_phase_diagram_database}, and for alloy thermodynamics with \gls{calphad}~\cite{lukas2007computational}. Over the past two decades, high-throughput computational approaches have further expanded the landscape with large-scale materials databases such as the Materials Project~\cite{jain2013commentary}, AFLOW~\cite{curtarolo2012aflow}, the Open Crystallographic Database~\cite{gravzulis2012crystallography}, NOMAD~\cite{draxl2018nomad}, JARVIS~\cite{choudhary2020joint}, Materials Commons~\cite{puchala2016materials}, and OQMD~\cite{kirklin2015open}, to name only a few. Efficiently finding and aggregating data across these different resources has recently become possible through the OPTIMADE platform~\cite{evans2024developments}.

Nevertheless, addressing a new scientific question through a \gls{ml}-based approach often requires generating new large, domain-specific datasets. On both experimental and computational fronts, this process can be time-consuming and costly, frequently relying on large infrastructures and complex workflows. This challenge has driven research groups to automate experimental procedures~\cite{tom2024self}. A recent survey highlights that \textit{accelerated discovery} or \textit{accelerated research} is often cited as the main motivation for laboratory automation and autonomy~\cite{hung2024autonomous}.

The emergence of \glspl{sdl}, also referred to as autonomous laboratories, represents a major step forward in experimental materials science, enabling accelerated materials design and discovery. These systems integrate multidisciplinary expertise combining robotic platforms, real-time characterization, machine learning, and automated decision-making to execute closed experimental loops with minimal human intervention~\cite{tom2024self,hung2024autonomous,JIANG2025100010,angelopoulos2024transforming,Ishizuki31122023,szymanski2023autonomous}.

More broadly, autonomous research extends beyond experimental laboratories and involves the integration of data infrastructure, knowledge representation, and adaptive learning systems across the entire scientific process. In parallel, computational materials science has increasingly adopted automated workflows, facilitated by workflow managers such as ASE~\cite{larsen2017atomic}, AiiDA~\cite{Huber2020,Uhrin2021}, PyIron~\cite{janssen2019pyiron}, Ewoks~\cite{de_nolf_ewoks_2024}, PsiFlow~\cite{vandenhaute2023machine}, Atomate2~\cite{ganose2025atomate2}, Snakemake~\cite{molder2025sustainable}, and Nextflow~\cite{di2017nextflow}, among others. These tools enable reproducible, large-scale, high-throughput simulations with full data provenance tracking. Computational workflow frameworks thus complement experimental autonomy by providing scalable, traceable pipelines for \gls{dft}, \gls{md}, and \gls{mlip} training, forming the digital backbone required for true closed-loop automation. This integration is often coupled with active learning approaches in various contexts~\cite{smith2021automated,jose2024informative}.
A recent example of such a development is the automated computational workflow built for alloy phase diagram prediction, which starts from electronic structure data and incorporates an active learning closed loop for the training and validation of \gls{mlip}~\cite{menon2024electrons}.

Beyond efficiency and new capabilities, reproducibility is a central expectation for both \glspl{sdl} and computational workflows~\cite{hung2024autonomous}.
One of the key goals of any materials-oriented numerical platform, such as the DIAMOND platform~\cite{Diamond} developed for the French DIADEM initiative~\cite{pepr_diadem}, is to equip the materials science community with advanced, user-friendly tools and techniques that ensure research is reproducible
across computing environments.
Although reproducibility is widely recognized as a fundamental principle of scientific research, its implementation in computational and workflow-based approaches remains challenging~\cite{antunes2024reproducibility}.
Achieving a balance between performance and user-friendliness is essential to maintain both accessibility and broad dissemination within (and beyond) the \gls{hpc} community.

The aim of the present work is to develop a general-purpose container solution that combines the Guix package manager~\cite{Courts2015} with Apptainer~\cite{kurtzer_singularity_all_versions}, the continuation of the open-source version of Singularity~\cite{kurtzer_singularity_original_paper}. This combination has been identified as an efficient and high-performance strategy~\cite{antunes2024reproducibility}, and has already proven its robustness in geophysics~\cite{Bilke2025} for \gls{hpc} environments~\cite{8855563} for instance, supporting both CPUs~\cite{Zhang2017} and GPUs~\cite{https://doi.org/10.48550/arxiv.2503.21033}. The objective is to ensure adaptability to various workflow managers. Reproducibility and performance are evaluated across multiple computing environments and workflow systems, including ASE, AiiDA, and Ewoks.
Specific use cases include the training of a \gls{mlip}, automated dataset construction using \textit{ab initio} \gls{md} with systematic trajectory analysis for validation purposes, and a full-cycle atomistic comparison between experiment, theory, and simulation.

\section{Container solutions}\label{sec2}

\subsection{Containerization: Apptainer}

The rapid introduction of \gls{ml} approaches has significantly enhanced research in materials science.
A strategic issue is to readily deploy these tools in diverse computing environments.
However, a major obstacle remains software installation and dependency management.
Installing complex software stacks, either locally or on shared computing infrastructure, often requires administration privileges and can lead to dependency conflicts when multiple libraries or compiler versions coexist.

Dependency resolution has long been a challenge in scientific computing~\cite{zakaria2022hpcdependencies}.
A robust solution is provided by containers, which enable software isolation through encapsulation of the full execution environment.
A container includes the software itself, its dependencies, and runtime libraries, ensuring consistent behavior across different machines.
Multiple containers can operate on the same host system while sharing only the hardware and the host kernel.
In contrast to \gls{vm}, containers reuse the host operating system kernel, which results in significantly lower overhead and near-native performance, at the cost of requiring kernel compatibility between host and container.
Containers therefore provide reproducible, portable environments suitable for computational materials workflows.
They can embed pre-configured dependencies and simulation parameters, ensuring consistent execution of \gls{ml} training, data pre-processing, or post-processing tasks across computing facilities without additional setup. 

Building, testing, and executing containers requires a container management system.
Several solutions exist, including Docker~\cite{merkel2014docker}, 
Podman~\cite{heon_2018_podman}, Singularity~\cite{kurtzer_singularity_all_versions}, and Linux Containers (LXC)~\cite{lxc_issue_4007}.
Docker remains the most widely adopted platform, designed primarily for standalone applications, web services, and software development environments where users have administrative access. Docker’s requirement of elevated privileges limits its applicability in shared computing clusters and \gls{hpc} environments.

Apptainer, formerly known as Singularity, was specifically developed to address these constraints.
It operates within the Linux user namespace, enabling container execution with user-level permissions while maintaining root-like behavior inside the container.
Apptainer also provides native support for hybrid MPI applications by interfacing directly with the host’s MPI installation, thereby achieving near-native interconnect performance.
Additionally, it offers built-in compatibility with both NVIDIA and AMD GPUs without requiring additional configuration.
These features make Apptainer particularly suitable for scientific workflows combining \gls{hpc} performance requirements with strict reproducibility and security constraints.
It is for these reasons that Apptainer was chosen as our containerization solution for the DIAMOND platform~\cite{Diamond} of DIADEM~\cite{pepr_diadem}.
\begin{figure}[ht]
  \centering
  \includegraphics[width=\textwidth]{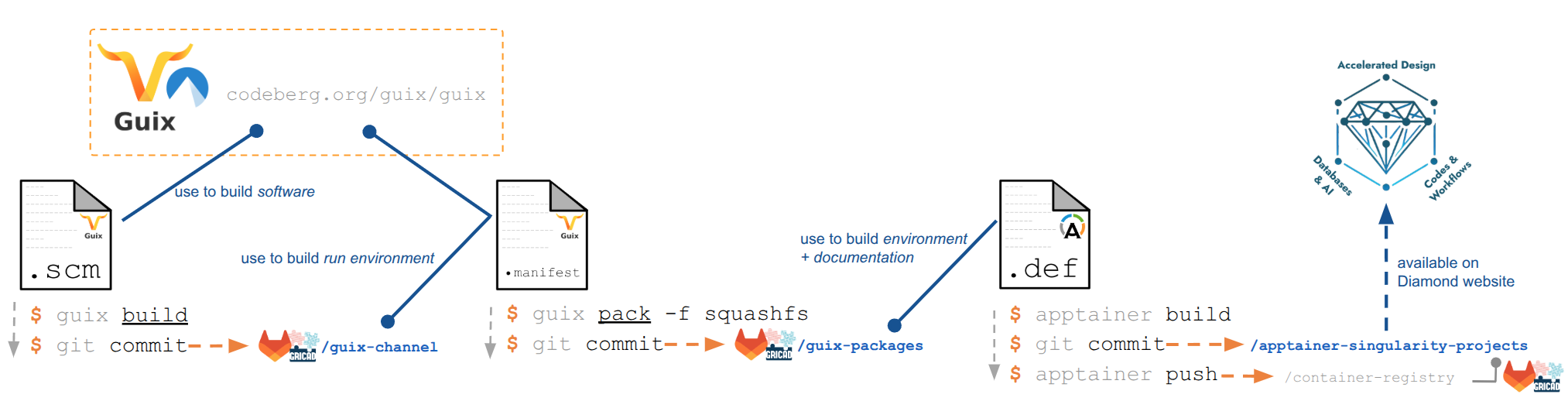}
  \caption{Schematic representation of the three repositories maintained and their interactions in the creation of fully reproducible container images. All repositories are publicly available on GRICAD's GitLab forge~\cite{gricad-gitlab}.}
  \label{fig:container-flowchart}
\end{figure}

\subsection{Package manager: Guix}

Container images usually weigh hundreds of megabytes (MB) to several gigabytes (GB), which makes storing multiple versions for long periods impractical. Definition files used to reconstruct these images are \textit{space-reproducible}, meaning that, at a fixed build time, the same output is produced on different machines, assuming they use the same Apptainer and kernel versions. The main limitation is that they are not \textit{time-reproducible}, because container definitions rely on operating-system or language-specific package managers, such as \texttt{apt} or \texttt{pip}. These tools are time-dependent: running the same \texttt{install} command at different moments may yield different versions of a package.

This time dependence means that container images built at different dates may provide different environments, which introduces uncontrolled variations in software installations. To reduce this issue, the package manager was replaced with GNU Guix, a functional package manager following the approach introduced by Nix~\cite{dolstra2004nix}. In Guix, each package is defined through a Scheme function, Scheme~\cite{dybvig2003scheme} being the language used to describe package builds. Inputs to these functions are the required dependencies, and outputs are the resulting binaries and libraries. This method represents a significant improvement compared to traditional deployment strategies, with strong guarantees on long-term reproducibility~\cite{Malka_2024}.

Lists of Scheme expressions form package “recipes.” Since these recipes are versioned using \texttt{git} --- where the upstream state corresponds to a specific commit of the Guix channel, Guix provides a mechanism to rebuild packages exactly as they were at that commit. This ensures bit-for-bit reproducible packages constructed with the same compilers and libraries originally used. Because exact reproducibility depends on the availability of the reference commit, ongoing collaborative efforts between Guix and the Software Heritage initiative~\cite{courtes2024source} further strengthen long-term software environment stability.

Guix allows users to create and manage software environments and to fix their time state using the \textit{time-machine} command, enabling both time- and space-reproducible environments~\cite{Vallet2022}.
Guix can also produce a \texttt{squashfs} archive from an environment. This archive is readable by Apptainer and contains all necessary libraries and binaries, providing a self-contained execution environment. The only additional requirement is minimal bash support (needed for the \texttt{apptainer shell} command), whose size ranges from about \qty{26}{\mega\byte} (minimal functionality) to \qty{37}{\mega\byte} (with user-friendly features and tab completion). When executed by a working Apptainer installation, these \texttt{squashfs} archives produce smaller, reproducible, task-specific containers, at the cost of preparing software with Guix rather than installing it through traditional OS package managers.

A fully integrated continuous development chain, based on three GitLab repositories, is shown in Figure~\ref{fig:container-flowchart}.
First, an open custom Guix channel stores and versions materials science software not available in the main Guix distribution.
Second, a private repository automatically builds \texttt{squashfs} archives using two inputs: a manifest listing the Guix packages to include, and a channel file specifying the exact commit (or commits, if multiple channels are used) from which the environment must be built.
Finally, a third open repository contains the Apptainer definition files. Most of them simply bootstrap the previously built archives and include container-specific documentation and configuration. A small number correspond to software that could not be packaged through Guix; these containers are less reproducible and larger, but they remain usable within workflows and can still be shared. This repository also provides a container registry allowing users to pull images directly from the platform.

\begin{figure}[t]
  \centering
  \includegraphics[width=0.98\textwidth]{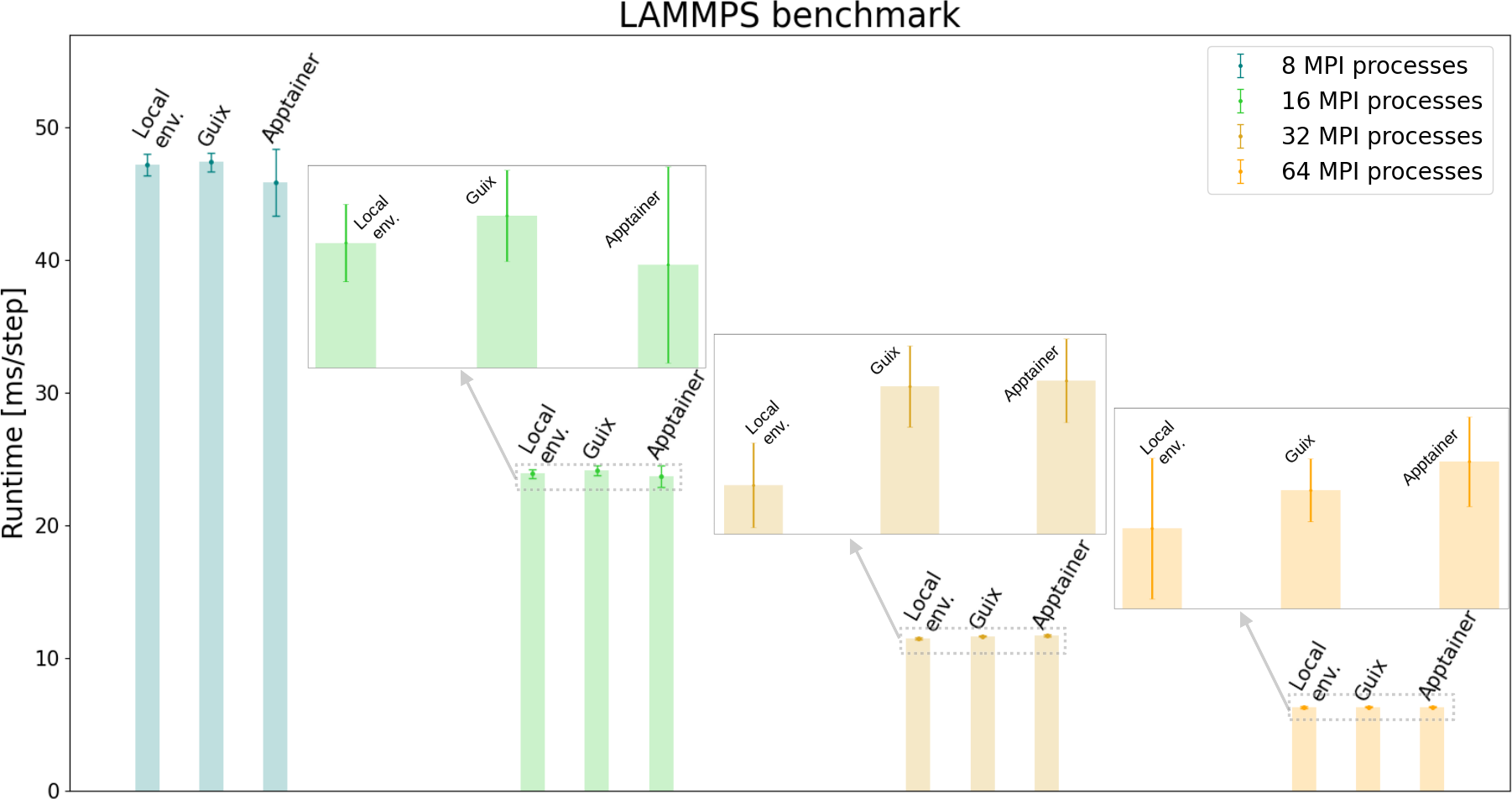}
  \caption{LAMMPS benchmark: Comparison of execution runtimes between the local installation, the Guix-packed version, and the Apptainer-containerized version of the DIAMOND platform~\cite{Diamond}.}
  \label{fig:appt-guix-lmp-bench}
\end{figure}
\subsection{Performance}

\subsubsection{General considerations}

Regarding the DIAMOND platform, one of the end goals of these software-environment technologies is their use in high-performance computing environments, making it important to verify that they allow efficient parallel execution of the software they contain. In theory, both Apptainer and Guix should introduce no computational overhead \cite{8855563}. In other words, for a given hardware configuration, the runtime should be the same whether the code runs natively on the host system or through these technologies.

For Apptainer, this expectation follows from the fact that containers use the host kernel directly and therefore do not require virtualization layers (unlike virtual machines). This appears to be true for pure computation tasks, but does not appears to be  exactly true when data needs to be shared periodically between the host and containers, mainly when the software writes and/or reads files \cite{8950978, Tondreau2025} thus explaining the difference that can appear between Guix-only and our Guix-based Apptainer images.

For Guix, the situation is even simpler: all software dependencies are stored on the host filesystem, and the Guix daemon creates symbolic links and adjusts environment variables (including \texttt{PATH}) to ensure correct loading at runtime.
In short, Guix changes only how dependencies are installed and organized, while program execution occurs on the host system as usual.

\subsubsection{Tests for the LAMMPS code}
The impact of these technologies on computational cost was first assessed using CPU-parallel benchmarks with the LAMMPS software \cite{thompson_lammps_2022}.
Microcanonical (NVE) molecular dynamics~\cite{allen_tildesley_2017} simulations were performed at room temperature on a \num{65536}-atom SiC supercell using a Modified Embedded Atom Model (MEAM) potential~\cite{kang2014governing}.
All tests were run on Intel Xeon 5218 machines of GRICAD’s \textit{Dahu} cluster~\cite{gricad_uga}.
The time step was set to \qty{1}{\femto\second}, and \num{1000} steps were carried out. Computations used $8$, $16$, $32$ or $64$ MPI processes.
The results in Figure~\ref{fig:appt-guix-lmp-bench} were averaged over $20$ simulations to reduce external variance; error bars indicate the standard deviation of the runtimes.

Figure~\ref{fig:appt-guix-lmp-bench} shows that no significant runtime variations or systematic overheads are observed. Increasing the number of MPI processes decreases runtime variability in a similar way for all execution modes, indicating that inter-process communication behaves similarly inside and outside containers.

\subsubsection{Tests for the training of MLIP}

A second performance test concerns the training of a potential using a High-Dimensional Neural Network  (HDNN)~\cite{behler2021four} with the N2P2 package~\cite{n2p2_CompPhysVienna_2025}. The training data correspond to liquid and crystalline boron structures extracted from \textit{ab initio} MD trajectories reported in a previous work~\cite{sandberg2023feature}. Benchmarks were performed using either one node with $1$, $2$, $4$, $8$, $10$, $16$, or $32$ MPI processes, or two nodes with $64$ processes (data not shown), on the same GRICAD \textit{Dahu} architecture.

Preliminary tests (not shown) indicate that N2P2 is mainly optimized for MPI-based distributed memory. OpenMP shared-memory parallelization appears to be unsupported or inefficient and was therefore not used here.

\begin{figure}[t]
  \centering
  \includegraphics[width=0.32\textwidth]{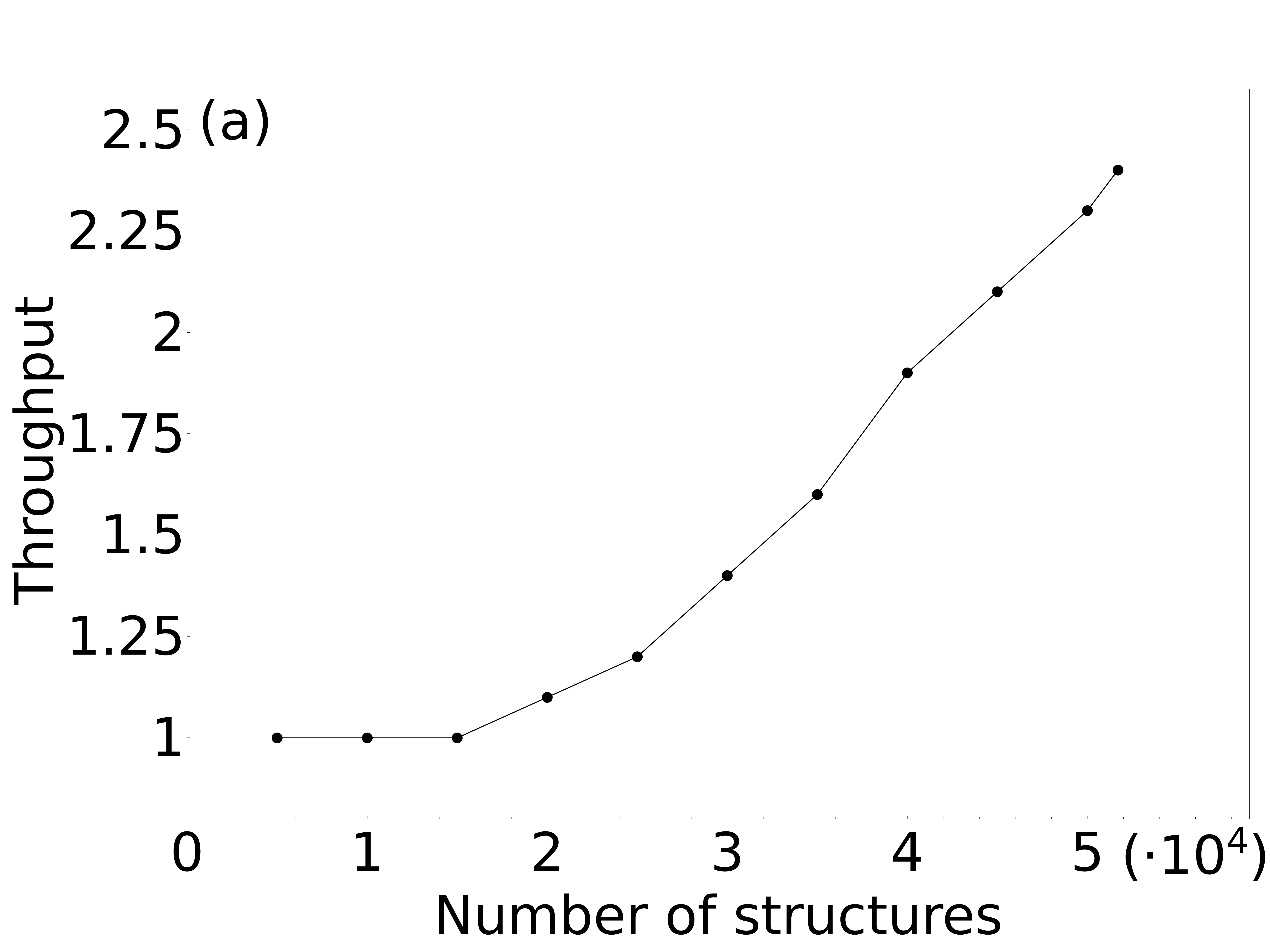}
  \includegraphics[width=0.32\textwidth]{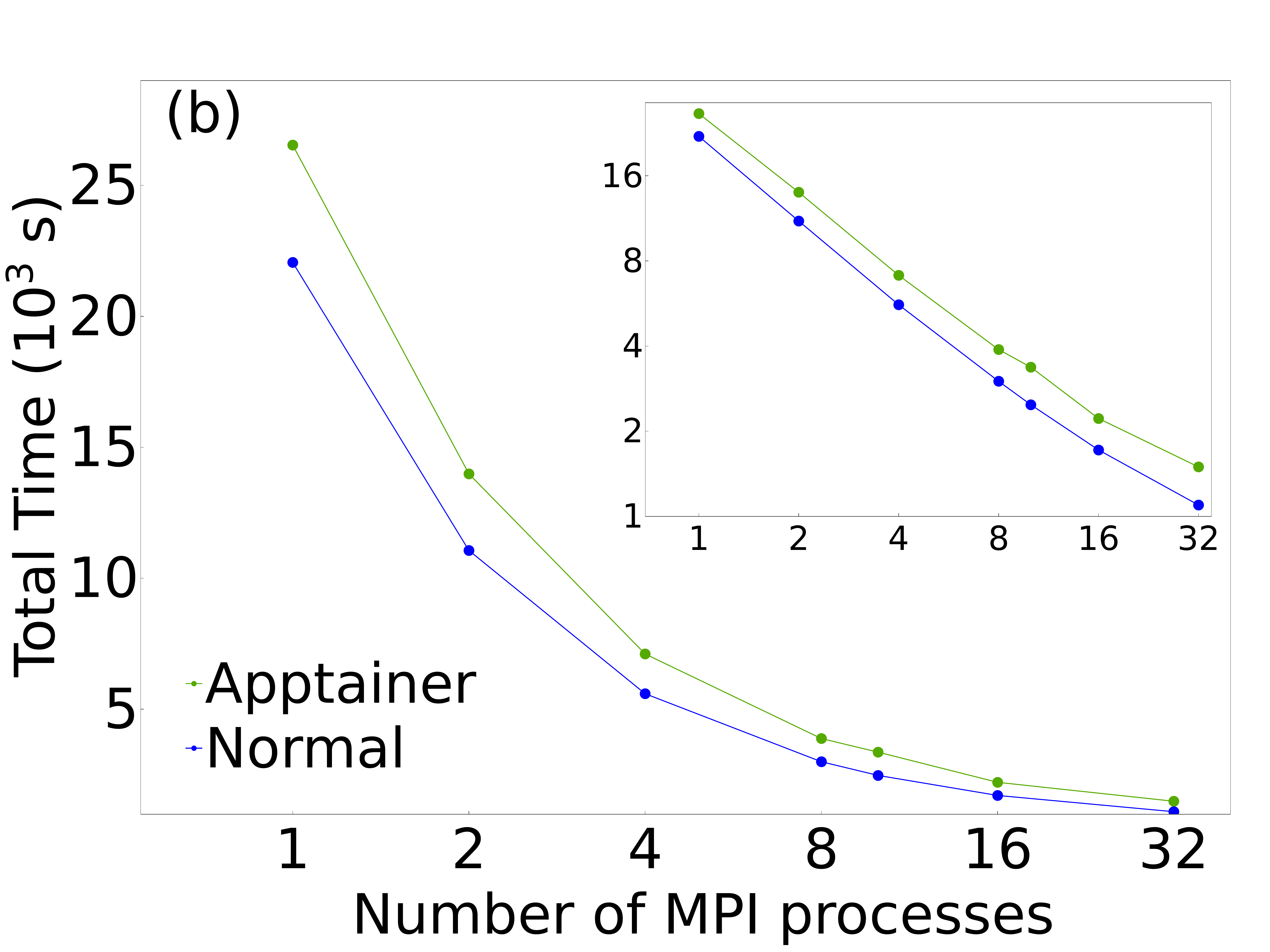}
  \includegraphics[width=0.32\textwidth]{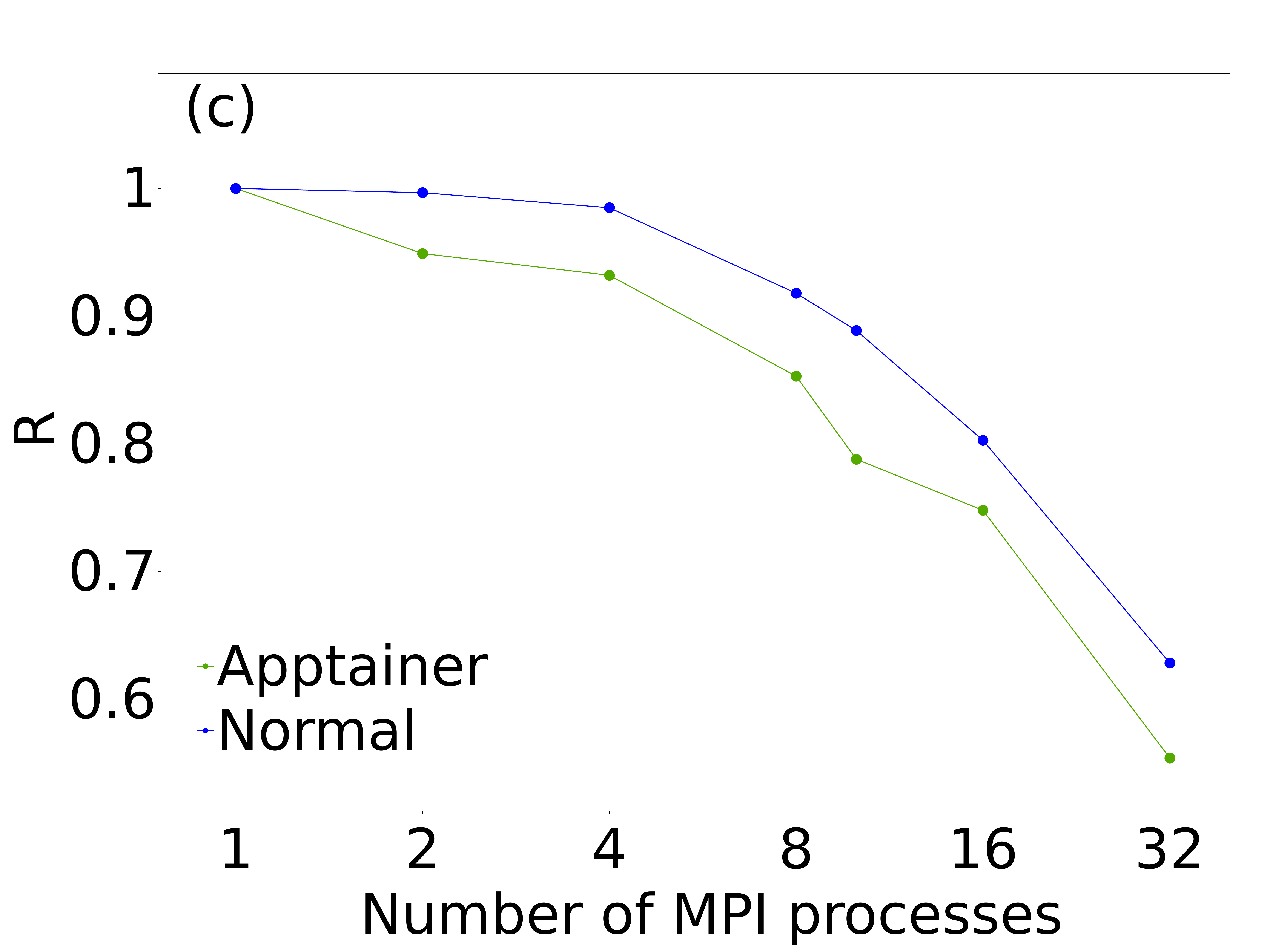}
  \caption{N2P2 benchmark:
    (a) Throughput of training runs as a function of dataset size for 32 MPI processes;
    (b) total time for 100 epochs using \num{51702} structures as a function of the number of MPI processes for the normal installation and the Guix-packed Apptainer containerization on the \textit{Dahu} machine;
    (b) absolute total time as a function of the number of MPI processes ($\log_2$--$\log_2$ scale) and corresponding efficiency $R$, plotted as in (c).
  }
  \label{fig:benchmark_n2p2}
\end{figure}

A first test examined the influence of increasing dataset size on training performance. Figure~\ref{fig:benchmark_n2p2}(a) shows the evolution of the throughput, defined as the total training time for $100$ epochs divided by the number of structures. Training was performed with dataset sizes from \num{5000} structures up to the full set of \num{51702} boron configurations~\cite{sandberg2023feature}, using $1$ node and $32$ MPI tasks. Dataset sizes range from \qty{69.4}{\mega\byte} (\num{5000} structures) to \qty{984.1}{\mega\byte} (\num{51702} structures).

This test provides insight into how N2P2 performance depends on dataset size. The throughput shows a clear decrease for larger datasets, consistent with the increasing I/O and computational effort.

A second test focused on the dependence of efficiency on the number of MPI processes. The number of training epochs was fixed to $100$, and runs were executed with the flag \verb|--bind-to-core| to ensure proper process binding and avoid overloading a single socket. Additional information on process binding and MPI execution is available in the official documentation~\cite{openmpi_mpirun_manpage}. The number of MPI processes was set to $n=$ $1$, $2$, $4$, $8$, $10$, $16$, and $32$. The comparison was made between the direct installation on \textit{Dahu}, so-called normal, and the Apptainer+Guix environment, since the previous benchmark indicated only minor differences between Guix alone and Apptainer+Guix.

Total training times were recorded, and the efficiency was computed as $R = t_1/(n,t_n)$, where $t_1$ is the runtime with $1$ MPI process and $t_n$ the runtime with $n$ processes. Figures~\ref{fig:benchmark_n2p2}(b) and (c) show the results. A visible overhead appears for the containerized execution at low process counts, and the efficiency $R$ is reduced across all process numbers. This behavior is largely explained by the substantial number of files generated at each training epoch by N2P2, leading to intensive disk access. Such I/O patterns have a stronger impact in containerized environments, independently of the number of MPI processes.

Regarding the optimal number of processes for the present training workload, the results indicate that increasing $n$ beyond $8$ provides diminishing returns. This is consistent with the drop in efficiency observed for larger process counts.

\section{Workflow solutions}\label{sec3}

\subsection{General concepts of workflow managers}

In the current era of AI and data-driven scientific discovery~\cite{Leng2023May}, the need for reproducible, automated, and scalable execution of data-intensive computations has led to the widespread use of computational workflows. Computational workflows provide a structured framework for executing multi-step processes that may involve several simulation codes, data pipelines, and analysis tasks. They manage these operations in a coordinated manner, handling concurrency, task ordering, and data dependencies to ensure reliable and efficient execution. A central concept in workflow systems is the distinction between control flow, which determines the ordering and branching of tasks, and data flow, which defines how intermediate outputs are produced and transferred between steps. This introduces an abstraction layer that hides much of the complexity of software interoperability and data handling. As a result, users can run advanced simulations and analyses without specialised technical skills, while benefiting from improved reusability, rapid prototyping, and more efficient knowledge transfer within research groups.

Workflows can be implemented at different levels of complexity. At the simplest level, researchers often rely on \textit{ad hoc} scripts or interactive notebooks, written in Bash, Python, or Jupyter. At a more advanced level, \glspl{wms} offer formal mechanisms to define, execute, and monitor workflows. These systems provide automated scheduling, provenance tracking, error recovery, and execution monitoring, enabling robust and scalable workflow management. In computational materials science, this distinction corresponds to the use of ASE for script-based \textit{ab initio} or atomistic calculations, and, at the other end, to \glspl{wms} such as AiiDA, Pyiron, and FireWorks~\cite{Jain2015Dec} for large-scale high-throughput studies requiring automation, provenance recording, and organised data handling.

A wide range of scientific \glspl{wms} has been developed in recent years, each providing mechanisms to store hyperparameters, workflow definitions, and the data used~\cite{Huber2020,Uhrin2021, janssen2019pyiron, de_nolf_ewoks_2024, ganose2025atomate2, Jain2015Dec, Mortensen2020}. However, the amount of execution information they capture varies significantly, leading to notable differences in transparency, efficiency, and reproducibility.
AiiDA is a representative \gls{wms} focused on rigorous reproducibility: each workflow execution generates a detailed provenance graph containing hyperparameters, task order, and all inputs and outputs. This information is stored in a relational database, where it can be queried to reconstruct the full computational history. Pyiron is another widely used \gls{wms}, providing an \gls{ide} for computational materials science. Based on Jupyter notebooks, it enables user-friendly and interactive development of simulation workflows.

Since no single \gls{wms} can meet all requirements, meta-workflow platforms such as Ewoks~\cite{de_nolf_ewoks_2024} have been developed. These systems can interface with multiple \glspl{wms}s and are particularly suited for experimental environments. Synchrotron facilities, for instance, face specific challenges due to a highly diverse user base, ranging from short-term visitors who need rapid onboarding to long-term groups developing instrumentation. Combined with the strict time constraints of beamtime, this leads to demanding data processing requirements, including the need for real-time analysis to guide experiments, support for intensive offline processing, and the ability to re-execute workflows. Ewoks addresses these needs by providing a unified interface to several execution engines, including Dask for distributed computing, Orange for graphical data workflows, and Pypushflow for control-loop execution.

\subsection{AiiDA workflow with VASP}


\subsubsection{Workflow scheme}
\begin{figure}[tb]
  \centering
  \includegraphics[width=0.9\linewidth]{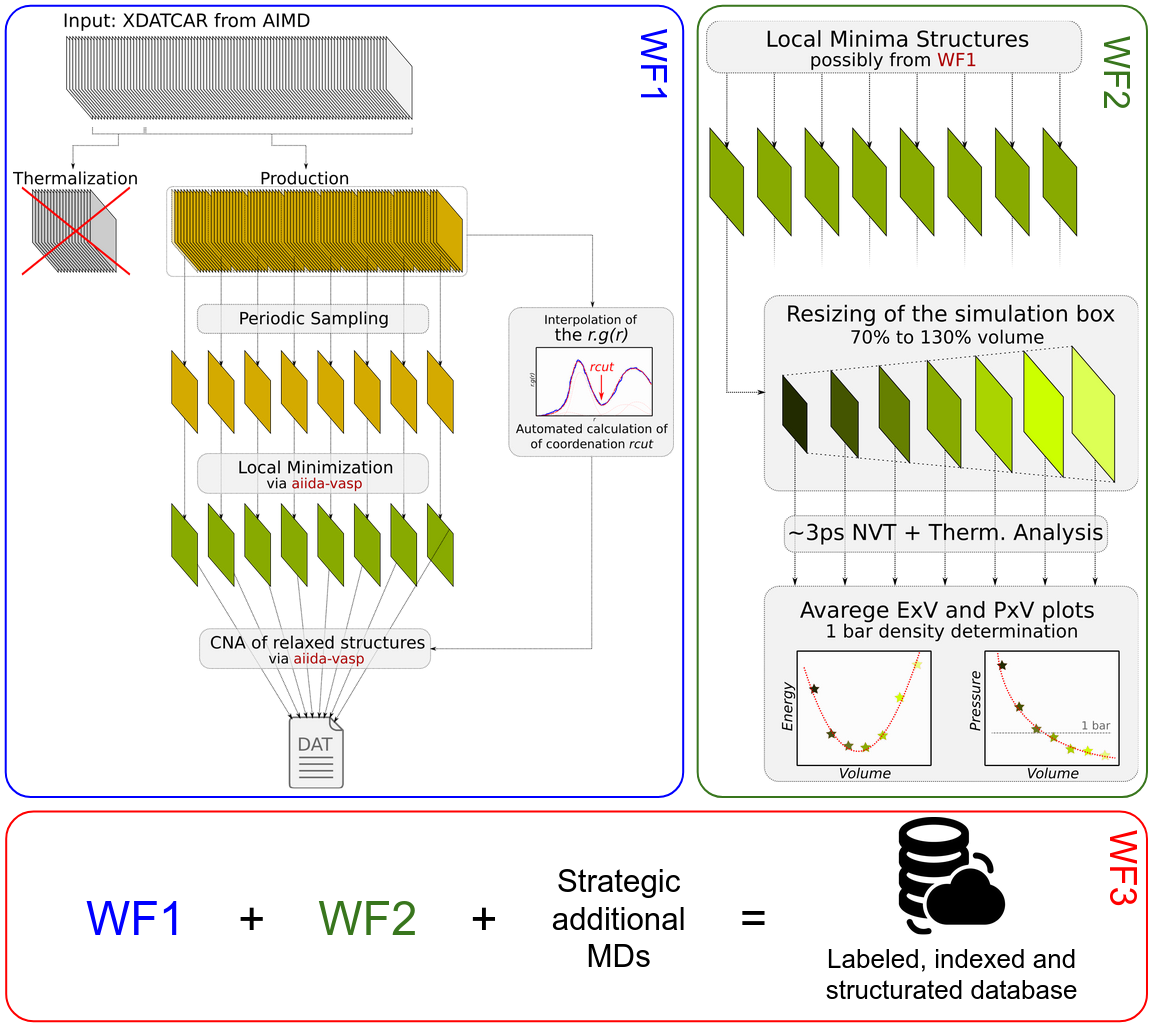}
  \caption{Schematic representation of the DIAMOND VASP workflow modules.
    (WF1) Inherent-structure extraction;
    (WF2) equilibrium volume estimation via EOS fitting;
    (WF3) integrated data-generation workflow for machine-learning force-field training.}
  \label{fig:vasp_md_workflow}
\end{figure}

There is a strong interest in the VASP code~\cite{Kresse1993,Kresse1996,Kresse1996_2} in the materials science community~\cite{hafner2008ab}, making it a natural choice for a first demonstrator of workflow solutions. To automate the generation and analysis of \textit{ab initio} \gls{md} (AIMD) data, a modular workflow was implemented within the AiiDA framework. This tool automate the preparation, execution, and post-processing of VASP-based simulations, enabling reproducible and scalable generation of AIMD ensembles suitable for high-throughput simulation analysis, database construction, and \gls{mlip} training. The workflow is organized into three modules, illustrated in Figure~\ref{fig:vasp_md_workflow}.

\paragraph{\textbf{WF1}:}
The first module processes an existing MD trajectory by reading atomic positions directly from the \texttt{XDATCAR} file. Representative configurations are selected at regular intervals after discarding the initial thermalization steps. Each selected configuration is structurally optimized to obtain the corresponding inherent structure, defined as the local minimum of the potential energy surface closest to the sampled liquid configuration. This characterization provides insight into the thermodynamic landscape of disordered systems. Common-neighbor analysis (CNA)~\cite{jakse2006local,faken1994systematic,POLAK2022110882} is carried out using a dedicated pipeline built on the free Python API of OVITO Basic~\cite{stukowski2010_ovito}. A system-specific cutoff radius $r_{cut}$ is used, determined automatically from the radial distribution function $g(r)$ of the initial trajectory. This module also enables systematic extraction of metastable configurations for property evaluation or later stages of data processing.

\begin{figure}[ht]
  \centering
  \includegraphics[width=1\linewidth]{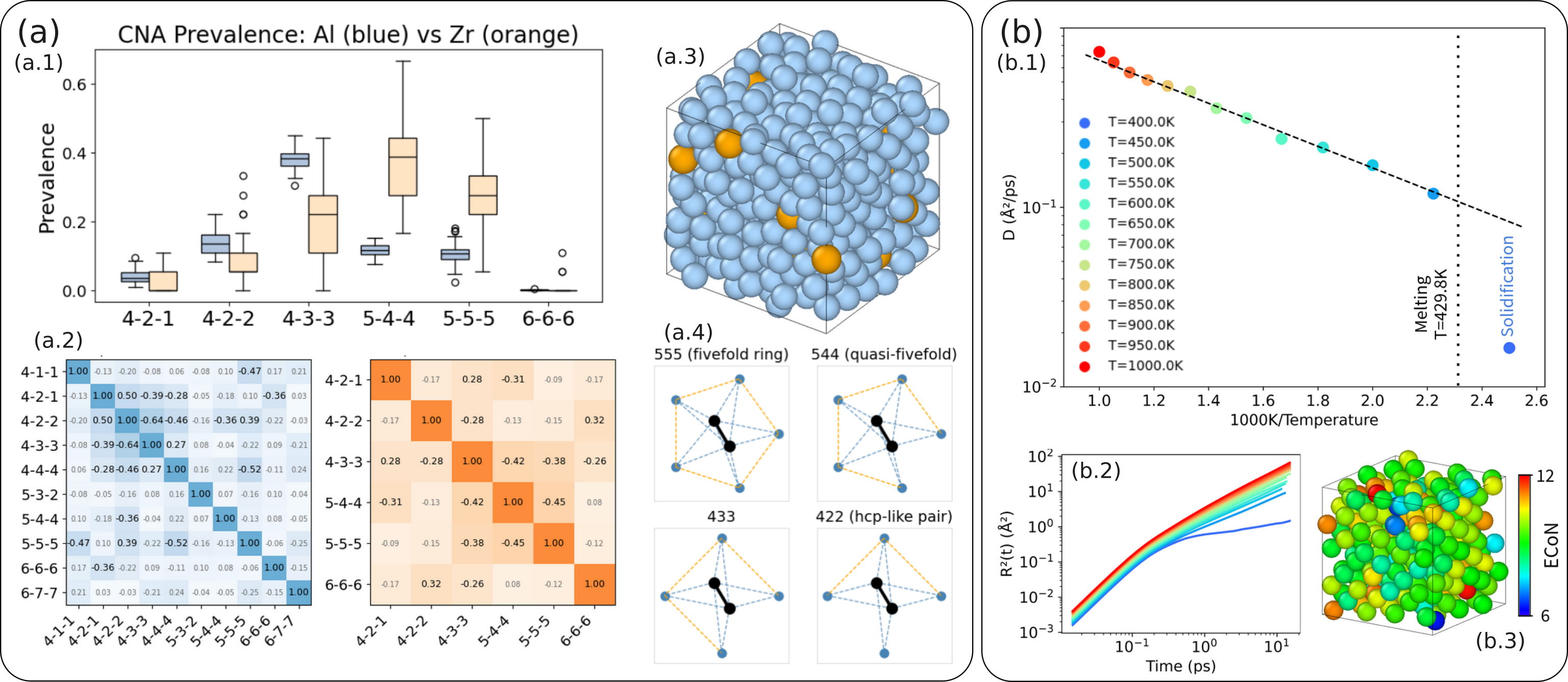}
  \caption{Example of results obtained with the automated VASP-based AiiDA workflow:
    (a) Statistical (a.1) and Spearman correlation (a.2) analyses of the local atomic arrangements in Al$_{97}$Zr$_{3}$ liquid alloys at $T = \qty{1500}{\kelvin}$ using WF1.
    The boxplot shows the distribution of the CNA signatures (a.4) for bonded-pair clusters such as 555, 544, 433, and 422~\cite{faken1994systematic}.
    The Spearman correlation matrices for Al and Zr indicate the dominant local ordering, consistent with the structural snapshot in (a.3).
    (b) Temperature dependence of the self-diffusion coefficient $D$ in liquid indium, shown as an Arrhenius plot (b.1).
    Diffusion coefficients are obtained from the slope of the mean-square displacement (b.2);
    a plateau in the MSD at the lowest temperature indicates partial crystallisation visible in the structure in (b.3), coloured as function of the Effective Coordination Number (ECoN)\cite{Hoppe1979}.}
  \label{fig:vasp_md_workflow_results}
\end{figure}
\paragraph{\textbf{WF2}:}
The second module takes as input a set of atomic configurations (for instance, files matching \texttt{/address/to/files/POSCAR*}) and performs a series of NVT \gls{md} simulations at different volumes. From the average pressure computed over the final portion of each short trajectory, the equilibrium volume is obtained by fitting the pressure--volume data to the third-order Birch--Murnaghan equation of state~\cite{birch1947finite}. This approach provides automated and reproducible estimation of the equilibrium density under fixed thermodynamic conditions and removes the need for manual trial-and-error cell adjustments. It is also significantly faster than performing a full NPT AIMD simulation to obtain the equilibrium volume.

\paragraph{\textbf{WF3}:}
The third module, currently under development, integrates WF1 and WF2 into a unified data-production pipeline for generating consistent datasets for property calculations or \gls{mlip} training and validation. This integration allows systematic sampling of the configurational space over a range of temperatures and pressures by combining long AIMD trajectories (WF1) with equilibrium density evaluation (WF2). High-pressure and solid-phase simulations are included to ensure that the datasets cover a broad region of the potential energy surface and therefore provide high-quality labeled data for force-field development within DIAMOND.

All modules use AiiDA’s provenance tracking and automated error-handling capabilities.
Job submission and monitoring on \gls{hpc} systems are handled via AiiDA’s transport layer, which supports several scheduling systems. Most simulations were executed on the GRICAD infrastructure using DIAMOND’s in-house plugin for interfacing AiiDA with clusters running the OAR scheduler. The workflow is designed in a flexible manner so that the \textit{ab initio} code can be replaced easily and additional analysis tools can be incorporated.

\subsubsection{Applications}

The AiiDA workflow with VASP~\cite{Haeuselmann2025aiida} was validated on two applications: (i) the structural analysis of Al--Zr liquid alloys from an AIMD simulation at $T = \qty{1500}{\kelvin}$ using WF1, and (ii) the temperature dependence of the self-diffusion coefficient in liquid indium (\ce{In}) over a wide range covering both the high-temperature liquid and the undercooled dense liquid, using a combination of WF1 and WF2. The corresponding results are shown in Figure~\ref{fig:vasp_md_workflow_results}.

Al--Zr alloys are promising candidates for additive manufacturing involving icosahedral mediated solidification (IMS)~\cite{pauzon2022novel}. Understanding the local arrangements induced by small Zr additions is essential but computationally demanding. Intensive AIMD simulations of Al$_{97}$Zr$_{3}$ with $500$ atoms were performed, and WF1 was used for high-throughput extraction, systematic optimization, error handling, and CNA analysis of 50 frames sampled from an AIMD trajectory at \qty{1500}{\kelvin}, which was necessary for meaningful statistics at low Zr concentration. Frank--Kasper polytetrahedral ordering around Zr atoms was identified, consistent with IMS~\cite{chapignac2025_icosahedral_AlZr}.

The search for universal diffusion behaviour in liquid metals raises fundamental questions about the underlying atomic mechanisms~\cite{jakse2016excess,demmel2021intimate,demmel2025diffusion}. Using WF1 and WF2, a systematic set of AIMD simulations with $250$ indium atoms was generated from the high-temperature liquid down to the undercooled regime in steps of \qty{50}{\kelvin}. A crossover from Arrhenius behaviour to a power law was observed near $1.8$ times the melting point~\cite{Almeida_2025_diffusion_indium}. Automated equilibrium volume determination at each temperature (performed efficiently using WF2) was essential for consistent diffusion coefficients and removed the need for manual volume adjustments that previously required significant user effort.

\subsection{Aiida workflow for path integral molecular dynamics}

We have identified emerging classes of hydrogen-containing materials with technologically relevant properties, in particular metal hydrides~\cite{Drozdov2019} and clathrates~\cite{FarrandoPerez2022}. To investigate the quantum motion of hydrogen nuclei, we developed a workflow based on the Path-Integral Molecular Dynamics (PIMD) formalism, including integration with MACE~\cite{Batatia2022mace,Batatia2022Design}  \gls{mlip} to accelerate simulations. An AiiDA plugin (\texttt{aiida-pioud}) was implemented to perform PIMD simulations using the PIOUD integration method~\cite{Morresi2021} in a scalable and reproducible manner~\cite{aiidapioudref}. PIMD captures nuclear quantum effects by mapping each quantum system onto a classical ring polymer in which each atom is represented by multiple replicas (beads) connected by harmonic springs in imaginary time.

\begin{figure}[t]
  \centering
  \includegraphics[width=0.415\textwidth]{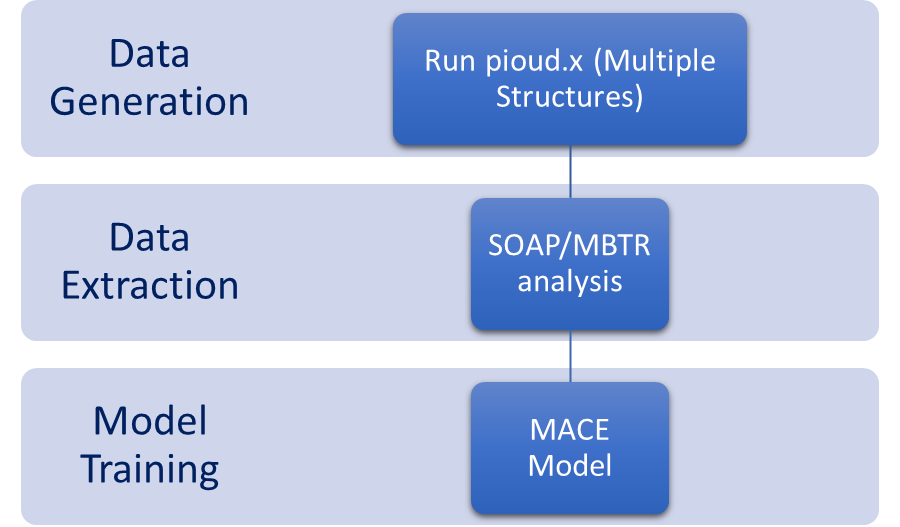}
  \includegraphics[width=0.45\textwidth]{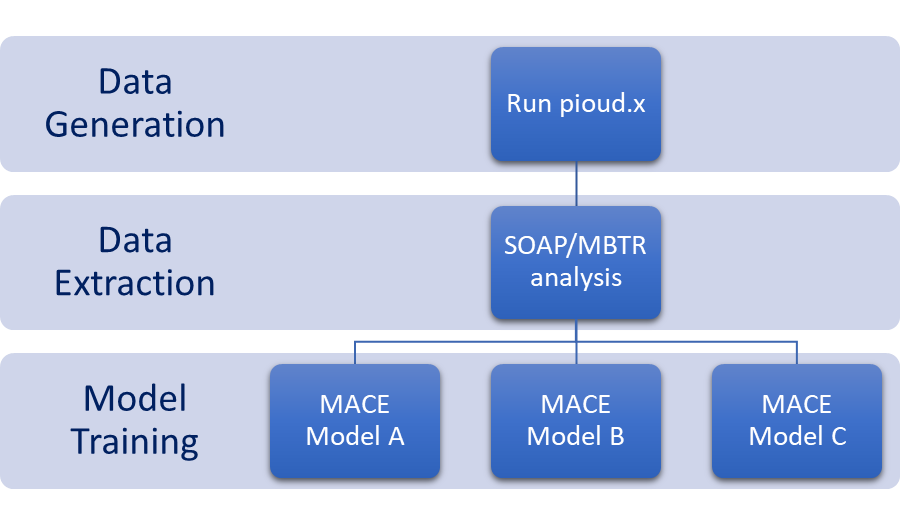}
  \caption{Left:~Workchain to obtain a single MACE model file from multiple \texttt{pioud.x} runs.
    Right:~Workchain to obtain multiple MACE model files for ensemble inference.\label{fig:aiidapioud}}
\end{figure}

The \texttt{aiida-pioud} plugin provides continuous workflow integration for the training of MACE \glspl{mlip} through AiiDA workchains and ensures reproducibility through AiiDA’s provenance tracking. Two workchains are currently available for generating MACE models. The first runs multiple PIOUD calculations on a set of input structures and aggregates the results. It then performs SOAP/MBTR-based structure selection~\cite{soap,Huo_2022,HIMANEN2020106949} using pioud\_process~\cite{pioudprocess}, a python library to identify distinct configurations for efficient MACE training, and finally produces a single MACE model file. The second workchain executes a single PIOUD calculation, applies SOAP/MBTR-based structure selection, and outputs multiple MACE models suitable for ensemble inference. A simplified representation of these workflows is shown in Figure~\ref{fig:aiidapioud}.

A containerised version of the associated codes (using Apptainer) is provided on the GRICAD GitLab server, enabling these calculations to be executed directly from JupyterLab notebooks.

Apart from this, an ISO-C binding wrapper was developed to enable access to MACE from any Fortran code~\cite{fortranmaceref}, allowing direct active-learning workflows. Integration of this wrapper into PIOUD is planned for the near future.

\subsection{Experimental and simulation workflow with Ewoks}\label{sec:ewoks}
In this section, FACETS (Full-circle Atomistic Comparison of Experiment, Theory and Simulation) is presented as a proof-of-concept Jupyter interface that uses Ewoks to automate a workflow developed at the ID01 beamline~\cite{leake_nanodiffraction_2019}.
This example illustrates the orchestration of on-demand, heterogeneous computations (using GPU, OpenMP or MPI) that uses separation of concerns by relying on Apptainer packages created from GNU Guix.

The methodology follows a multi-stage computational workflow to generate and validate an atomic-scale model from experimental \gls{bcdi}~\cite{robinson_coherent_2009} data. It consists in the following stages:

\begin{enumerate}
  \item Experimental Data Reconstruction: The workflow starts with raw \gls{bcdi} data, which is computationally reconstructed using a dedicated pipeline like cdiutils~\cite{atlan_cdiutils_2025} or bcdi~\cite{carnis_bcdi_2023} to produce a three-dimensional model of a nanoparticle electron density in an orthogonal reference frame.
  \item Geometric and Crystallographic Analysis: The reconstructed volume is analyzed to identify its external facets (\textit{i.e.}, the types of \{hkl\} crystallographic facets) and determine the overall crystallographic orientation. This analysis provides the necessary parameters for the subsequent modeling steps and also yields direct experimental insights into the nanoparticle’s morphology and structure.
  \item Atomic Model Generation and Refinement: The extracted orientation and a mesh of the particle's surface are fed into an OpenMP version of the nanoSCULPT software~\cite{prakash_nanosculpt_2016} to generate an initial, unrelaxed atomic structure. To obtain a physically realistic configuration, this structure is then minimized using a LAMMPS~\cite{thompson_lammps_2022} geometry optimization, resulting in a final set of minimized atomic positions.
  \item Validation and Comparison: Finally, the simulated atomic structure undergoes the same facet analysis as the initial experimental reconstruction. This allows for a quantitative comparison between the final simulated model and the experimental data (\textit{i.e.}, the recovered lattice displacement and strain), validating the accuracy of the generated atomic structure.
\end{enumerate}

\begin{figure}[tb]
  \centering
  \includegraphics[width=0.7\textwidth]{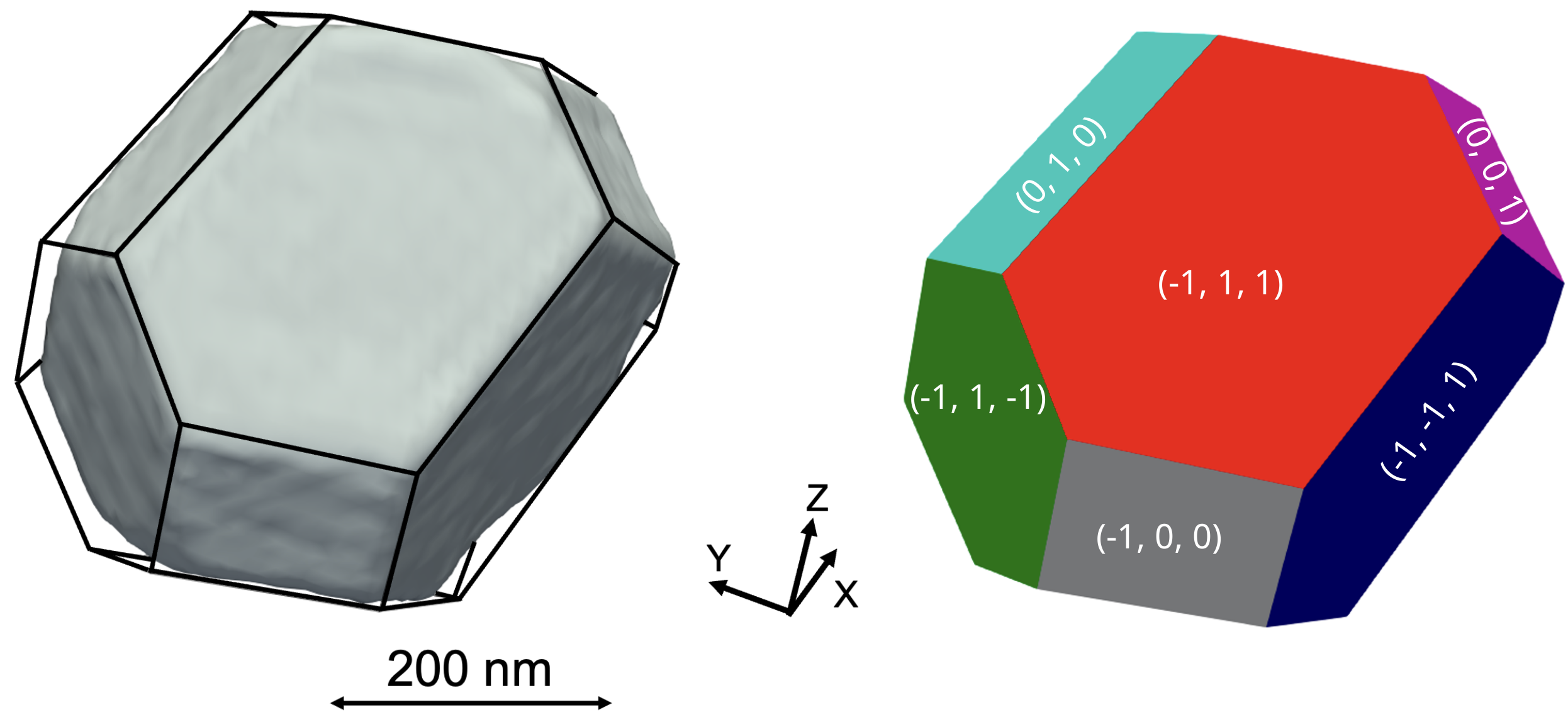}
  \caption{Left:~Example of a 3D electron density reconstruction of a platinum nanoparticle derived from a \gls{bcdi} dataset.
    Right:~Crystallographic facet indexing of the same particle using the FACETS workflow, where the white labels $(h,k,l)$ correspond to the Miller indices denoting the orientation of each surface plane.\label{fig:nanoparticles}}
\end{figure}

Figure~\ref{fig:nanoparticles} presents an example of facet indexation obtained with the second stage of the FACETS workflow, applied to a three-dimensional electron density reconstruction of a platinum nanoparticle derived from \gls{bcdi}.

Performance benchmarks on that same nanoparticle with \num{48}~cores demonstrate the workflow's adaptability for real-time applications.
Stages~1 and~2 represent a relatively fixed computational cost: Stage~1 typically requires \qtyrange{1}{3}{minutes}, varying slightly due to the nature of reconstruction algorithms, while Stage~2 completes in under one minute.
The computational cost of Stages~3 and~4 scales with the desired resolution, \textit{i.e.}, the number of atoms.
As shown in Table~\ref{tab:timings}, reducing the scale factor decreases execution time.
This is sensible, as the simulated results remain qualitatively consistent across these scales~\cite{atlan_imaging_2023}.
This allows users to utilize lower-scale simulations for rapid, near real-time feedback during experiments, while reserving higher-scale simulations for intensive offline analysis.

This work transformed a mostly manual pipeline that took a full working day for the scale~$1/10$ into an automatic workflow than can run in less than ten minutes.
Furthermore, the reliance on Apptainer and GNU Guix ensures that this Jupyter workflow can be easily ported to other infrastructures, including personal machines, in a reproducible manner.

\begin{table}[t]
  \centering
  \caption{Execution times for Stages~3 and~4 of the FACETS workflow. The imposed relative force tolerance in the LAMMPS code is $10^{-12}$.\label{tab:timings}}
  \begin{tabular}{c S[table-format=1.1e1] S[table-format=<2.0] S[table-format=4.0] S[table-format=3.0]}
    \toprule
    \multirow{2}{*}{\textbf{Scale}} & \multicolumn{1}{c}{\multirow{2}{*}{\textbf{Atoms}}} & \multicolumn{2}{c}{\textbf{Stage 3} (\unit{\s})} & \multicolumn{1}{c}{\multirow{2}{*}{\textbf{Stage 4} (\unit{\s})}}       \\
    \cmidrule(lr){3-4}
                                    &                                                     & {nanoSCULPT}                                     & {LAMMPS}                                                          &     \\
    \midrule
    $1/20$                          & 2.4e5                                               & <1                                               & 24                                                                & 32  \\
    $1/10$                          & 1.9e7                                               & 3                                                & 255                                                               & 140 \\
    $1/5$                           & 1.5e8                                               & 22                                               & 5221                                                              & 886 \\
    \bottomrule
  \end{tabular}
\end{table}

\section{Conclusion}

This work presented a reproducible and portable computing solution that addresses a long-standing practical difficulty in materials science: maintaining complex software stacks and workflows in a usable and reliable form over time. By combining the GNU Guix functional package manager with the Apptainer container system, it was shown that environments can be constructed that are reproducible in a strict, long-term sense while remaining sufficiently performant for deployment on modern HPC architectures without measurable overhead. The benchmarks confirm this: for standard CPU-parallel workloads, execution inside a container is essentially indistinguishable from native runs, and even I/O-intensive applications exhibit predictable and controlled behaviour.

An important outcome of this study is that such an environment does not only simplify software management but also supports new scientific practices. The integration with workflow managers such as AiiDA and Ewoks illustrates this. In the AiiDA case, automated VASP and Mace with path-integral molecular dynamics workflows, can be executed with full provenance tracking and without extensive manual supervision. The Ewoks-based FACETS workflow further demonstrates the ability to coordinate heterogeneous tasks during synchrotron experiments, reducing procedures that previously required hours to timescales compatible with real-time feedback.

Together, these developments contribute to the digital infrastructure required for autonomous and data-driven materials research, a central objective of the DIAMOND platform within the DIADEM program. Beyond reproducibility, the Guix--Apptainer approach lowers the entry barrier for new users, facilitates collaboration across groups, and ensures that datasets, workflows, and simulation environments can be shared and reused over long periods. In a context where scientific results increasingly depend on complex software layers, this level of clarity and stability is essential.

Future work will focus on extending the set of materials-science codes packaged with Guix, improving GPU support, and refining active-learning workflows for \gls{mlip} development. Even in its current state, the platform constitutes a significant step toward a unified, sustainable, and community-driven ecosystem for computational and experimental materials research.

\medskip
\textbf{Supporting Information} \par 
Supporting Information is available from the Wiley Online Library or from the author.

\medskip
\textbf{Acknowledgements} \par 
ÉP thanks Clément Atlan, Corentin Chatelier and Wout de Nolf for useful discussions and comments.
This work was supported by the French government “France 2030” initiative, under the DIADEM program managed by the “Agence Nationale de la Recherche” (ANR-22-PEXD-0015, DIAMOND).
We acknowledge the CINES and IDRIS under Project No. INP2227/72914, as well as CIMENT/GRICAD for computational resources. We acknowledge financial support under the French--German project PRCI ANR-DFG SOLIMAT (ANR-22-CE92-0079-01).
This work has benefited from a French government grant managed by the Agence Nationale de la Recherche under the France 2030 program, reference ANR-23-IACL-0006.

\medskip

%

\textbf{References}\\
\bibliographystyle{MSP}
\bibliography{references}

\end{document}